\documentclass{PoS}

\usepackage{graphicx}
\usepackage{amsmath}
\usepackage{amssymb}

\newcommand{\be}{\begin{equation}}
\newcommand{\ee}{\end{equation}}
\newcommand{\bea}{\begin{eqnarray}}
\newcommand{\eea}{\end{eqnarray}}
\newcommand{\bi}{\begin{itemize}}
\newcommand{\ei}{\end{itemize}}
\newcommand{\ben}{\begin{enumerate}}
\newcommand{\een}{\end{enumerate}}
\newcommand{\bt}{\begin{tabbing}}
\newcommand{\et}{\end{tabbing}}

\title{
   \begin{picture}(0,0)(0,0)%
   \end{picture}%
Finite size scaling for 4-flavor QCD
with finite chemical potential
}

\ShortTitle{
Finite size scaling for 4-flavor QCD
with finite chemical potential
}

\author{
   \speaker{Shinji Takeda}$^{a,b}$\thanks{E-mail: takeda@hep.s.kanazawa-u.ac.jp}, 
   Xiao-Yong Jin$^b$,
   Yoshinobu Kuramashi$^{b,c,d}$,
   Yoshifumi Nakamura$^b$,
   and
   Akira Ukawa$^d$
   \\
   \\
   \\
   \llap{$^a$}
Institute of Physics, Kanazawa University, Kanazawa 920-1192, Japan
   \\
   \llap{$^b$}
RIKEN Advanced Institute for Computational Science,
Kobe, Hyogo 650-0047, Japan
   \\ 
   \llap{$^c$}
Graduate School of Pure and Applied Sciences,
University of Tsukuba,
Tsukuba, Ibaraki 305-8571, Japan
   \\
   \llap{$^d$}
Center for Computational Sciences,
University of Tsukuba,
Tsukuba, Ibaraki 305-8577, Japan
}
   
\abstract{
We explore the phase diagram spanned by the temperature and the chemical potential for 4-flavor QCD by the phase-reweighting approach.
In order to determine the order of phase transition, we perform finite size scaling studies for various quantities, for example, susceptibility, kurtosis and Challa-Landau-Binder cumulant.
At the parameter ($\beta=1.60$, $\kappa=0.1371$, $c_{\rm sw}=1.9655$ and $N_{\rm T}=4$),  where the Kentucky group reported a first-order phase transition in their canonical simulation, we observe that the transition is consistent with being of first order.
}

\FullConference{The 30th International Symposium on Lattice Field Theory\\
		 June 24 -- 29,  2012\\
		 Cairns, Australia}

\begin{document}

\section{Introduction}

\begin{figure}[b]
\begin{center}
\scalebox{0.22}{\includegraphics{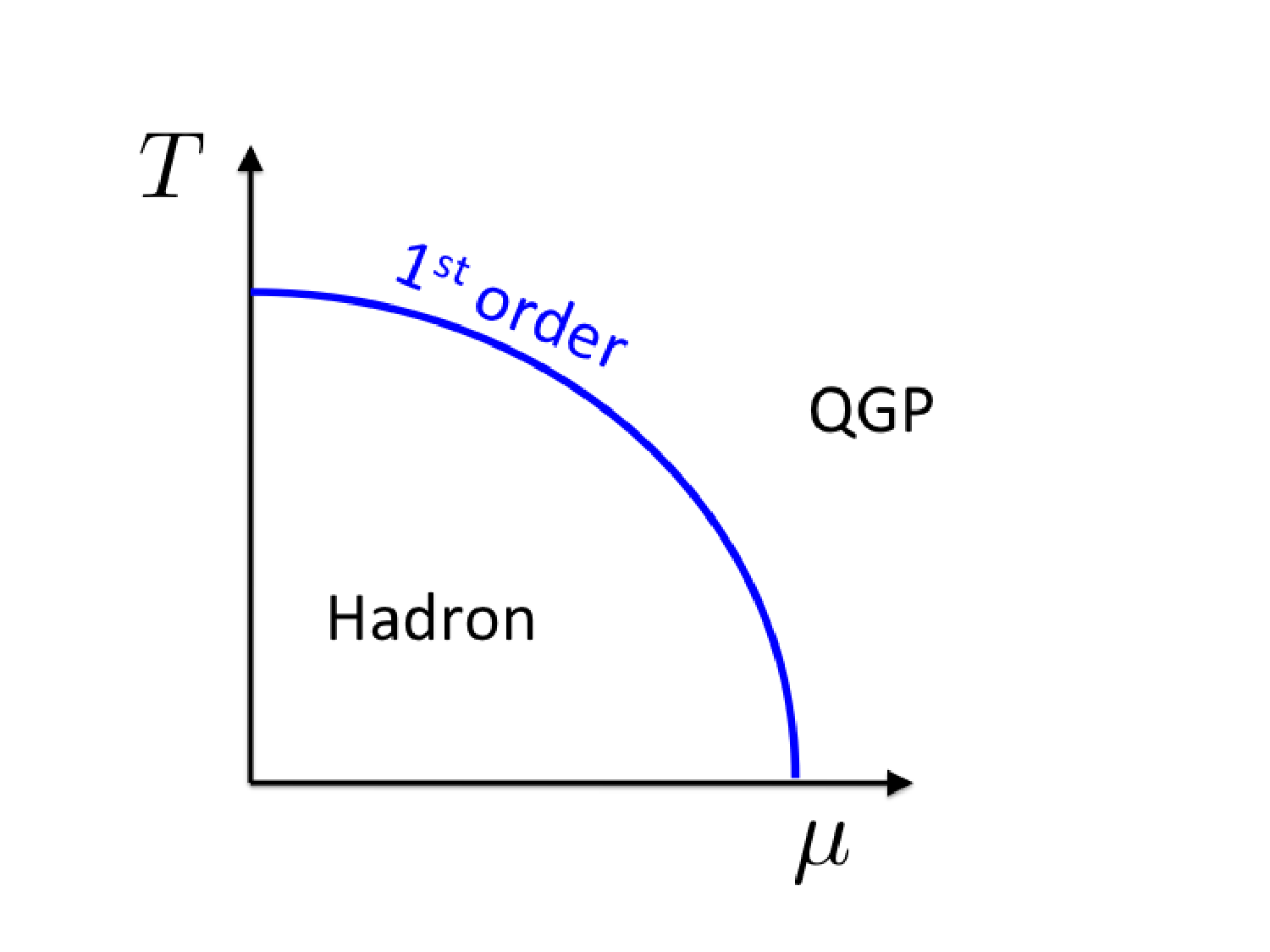}}
\end{center}
\caption{An expected phase diagram for 4-flavor QCD.
}
\label{fig:phasediagram}
\end{figure}

As is well known, the  4-flavor QCD is a good testing ground with finite temperature and chemical potential before studying the physically more relevant case of the 3-flavor theory.
One of the physical reason\footnote{
Another reason is the practical one that 4-flavor theory is suited for staggered fermions.
New ideas to attack finite density QCD, for example \cite{Fodor:2001au,D'Elia:2002gd,Kratochvila:2005mk}, were firstly examined in this theory.
}
for this is that the phase diagram of the former theory is expected to have a first-order phase transition line as shown in Fig.\ref{fig:phasediagram}.
It is empirically known that the first-order phase transition persists even for relatively large quark masses before turning into a crossover.
Therefore one should be able to detect the transition line with a reasonable computational cost and learn the physical characteristics of the transition.
The Kentucky group \cite{Li:2010qf} studied the phase structure of the 4-flavor theory using the canonical approach with Wilson-type fermions.
They observed an S-shape structure in the chemical potential vs. quark number plot which they took to be an indication of a first-order phase transition.
The study was made only for single lattice volume $6^3\times4$, however,  so this is not a conclusive statement.

Our main purpose here is to carry out a finite size scaling study for the 4-flavor QCD with finite chemical potential and to learn how we can quantitatively decide the order of transition within the grand canonical approach.
In the 3-flavor theory, especially for the physical spectrum of the up ,down and strange quarks, it is expected that the transition is weak, being a crossover at zero chemical potential and turning into a first-order one with an increasing chemical potential, and it would be hard to decide the order of transition.  
Since the transition in 4-flavor theory is also expected to be weak for heavy quark masses, we should be able to gain useful experience before tackling a more difficult 3-flavor  theory.

\section{Simulation setup}
We use the grand canonical approach.  The partition function is given by
\be
{\cal Z}_{\rm QCD}(\mu)
=
\int [dU]e^{-S_{\rm G}}[\det D(\mu)]^4.
\ee
We employ the Wilson-clover fermion action setting the number of flavors to be four.
The reason why we adopt the Wilson type fermions is that in future we want to smoothly move to the 3-flavor case.
In order to overcome the complex Boltzmann factor, we employ the phase-reweighting method.
Our approach may seem too straightforward since the phase-reweighting is usually used to explain how the sign problem appears and why it is so difficult to solve.
However, in \cite{Takeda:2011vd},  we learned that to some extent the phase can be controlled 
by increasing the temporal lattice size while keeping the other parameters fixed.
Therefore there may be a chance that this method provides useful information.

In the actual simulations we generate configurations by the phase-quenched partition function with $\mu\neq0$
\be
{\cal Z}_{||}(\mu)
=
\int [dU]e^{-S_{\rm G}}|\det D(\mu)|^4,
\ee
where the absolute value of the determinant ($\det D=|\det D|e^{i\theta}$) is taken into the Boltzmann weight.
Then after measuring observables and correcting the phase factor, we obtain the target results.
\be
\langle
{\cal O}
\rangle
=
\frac{\langle{\cal O}e^{i4\theta}\rangle_{||}}{\langle e^{i4\theta}\rangle_{||}}.
\ee
In the present work, we do not employ any parameter reweighting such as the 
$\beta$-reweighting or $\mu$-reweighting.

A key point in this  kind of calculation is how to obtain the phase.
We use the reduction technique proposed in Ref.\cite{Danzer:2008xs} and obtain the phase as well as quark number exactly,
that is, we compute them without introducing any systematic errors.
Actually, computing the phase is the hot spot of the calculation, and using GPU helps us to accumulate enough statistics.
Details of the calculation will be given in a future publication.

The lattice action which we use is the combination of Wilson-clover fermion action and Iwasaki gauge action.
The parameters are the same as those of the Kentucky group \cite{Li:2010qf},
$\beta=1.6$, $\kappa=0.1371$, $c_{\rm sw}=1.9655$ and $N_{\rm T}=4$.
This parameter set corresponds to $a=0.33$fm, $m_\pi=830$MeV and $T=150$MeV.
In contrast to the canonical approach by the Kentucky group, since we are using the grand canonical approach, the control parameter for quark number is now the chemical potential.  We cover the range $a\mu=0.1-0.35$ in our simulation.
Readers may wonder that the phase-reweighting method breaks down due to a charged pion condensate.
Since the condensate is expected to start from the value $a\mu_c(T=0)=am_\pi/2=0.7$, however, 
we do not need to worry about it.
To carry out  finite size scaling analyses, we chose 4 spatial volumes,
$6^3$, $6^2\times8$, $6\times8^2$ and $8^3$, while the temporal lattice size is fixed at $N_{\rm T}=4$.

\section{Simulation results}

\begin{figure}[t]
\begin{center}
\scalebox{0.9}{\includegraphics{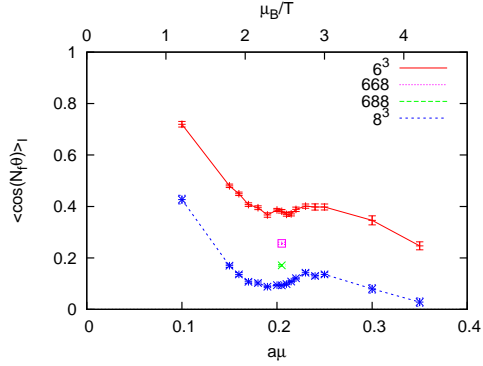}}
\end{center}
\caption{The phase-reweighting factor with $N_{\rm f}=4$ as a function of the chemical potential
(the baryon chemical potential in temperature unit is shown in the upper x-axis).
}
\label{fig:phasereweighting}
\end{figure}

In Fig.\ref{fig:phasereweighting} we show the phase-quenched average of the reweighting factor as a function of $a\mu$.  
For larger volume and $\mu$, the average reweighting factor tends closer to zero, so the sign problem is becoming more serious as expected.  However, since the reweighting factor remains non-vanishing beyond statistical errors, the  phase-quenched average is under control for the lattice volumes and the parameter sets used in the present simulations. 

An interesting observation is that there is a local minimum around $a\mu=0.2$.
We expect this to be the region of transition, which 
can be understood as follows.
Remembering the definition of the reweighting factor
$\langle e^{i4\theta}\rangle_{||}={\cal Z}_{\rm QCD}/{\cal Z}_{||}$, we observe that 
a zero of the average reweighting factor is related with a zero of the QCD partition function,  which is a transition point of the theory.
Of course, we do not expect a real phase transition but only a remnant for finite volumes, and hence no actual zeros for the real parameter space but just a remnant, such as a minimum.
The dip in Fig.\ref{fig:phasereweighting} may be  considered to be such a remnant, in which case the parameter region around $a\mu=0.2$ is expected to contain the transition point.

\begin{figure}[b]
\begin{center}
\begin{tabular}{cc}
\scalebox{0.77}{\includegraphics{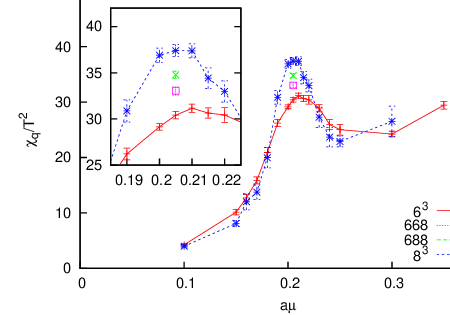}}
&
\hspace{10mm}
\scalebox{0.87}{\includegraphics{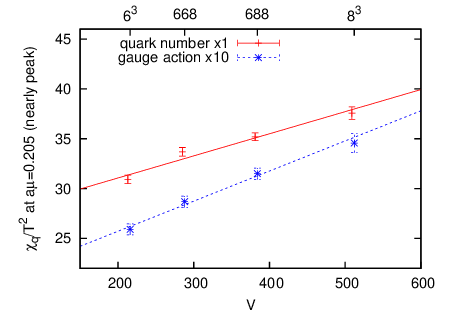}}
\end{tabular}
\end{center}
\caption{Susceptibility of quark number as function of $a\mu$ (Left) and
the volume scaling of peak height of the susceptibility (Right).
The curves in the right figure is a linear fit of the volume $\alpha+\beta V$
where $\alpha$ and $\beta$ are fitting parameter.
Note that in the right figure the value is evaluated at $a\mu=0.205$
which is nearly a peak position.
}
\label{fig:qnum_sus}
\end{figure}

Next, let us see the susceptibility of quark number shown in the left panel in Fig.\ref{fig:qnum_sus}.
Here we observe a clear volume dependence and the peak grows for larger volume.
Therefore it is likely that there is a phase transition.
The next question is what is the order of the phase transition.
To answer this question, let us see the volume dependence of the peak as shown in the right panel in Fig.\ref{fig:qnum_sus}.
We observe a linear volume dependence.
Other physical quantities including the gauge action also shows a similar behavior.
The volume dependence of the susceptibilities is consistent with a first-order phase transition.

\begin{figure}[t]
\begin{center}
\scalebox{0.75}{\includegraphics{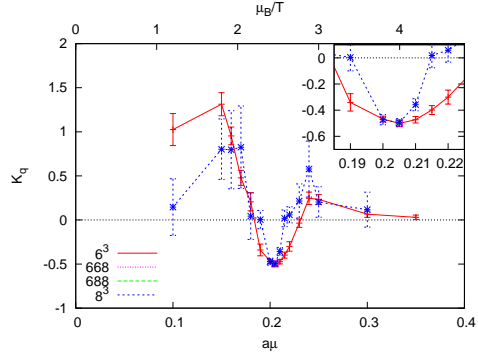}}
\end{center}
\caption{Kurtosis of quark number as function of $a\mu$.
}
\label{fig:qnum_krt}
\end{figure}

The kurtosis is defined from the fourth and second derivatives by 
\be
K_{\rm q}
=
\frac
{(\ln{\cal Z}_{\rm QCD}(\mu))^{\prime\prime\prime\prime}}
{((\ln{\cal Z}_{\rm QCD}(\mu))^{\prime\prime})^2}. 
\ee
Note that this quantity is slightly different from the 4-th order Binder cumulant $B_4=K+3$.
The results are plotted  in Fig.\ref{fig:qnum_krt} showing a negative value and a minimum 
around $a\mu=0.2$.
In the thermodynamic limit the kurtosis is expected to behave as 
\bea
\lim_{V\rightarrow\infty}K_{\rm q}=-2&:&\mbox{1st order}\\
-2<\lim_{V\rightarrow\infty}K_{\rm q}<0&:&\mbox{2nd order}\\
\lim_{V\rightarrow\infty}K_{\rm q}=0&:&\mbox{cross over}
\eea
As you can see from the figure the volume dependence of the minimum is very small and furthermore, the minimum value is far from $-2$.
Therefore, it is hard to conclude that the teansition is of first order from kurtosis. 
We discuss the discrepancy in the volume dependence of the susceptibility and the kurtosis in the next section.

\begin{figure}[t]
\begin{center}
\begin{tabular}{cc}
\scalebox{0.9}{\includegraphics{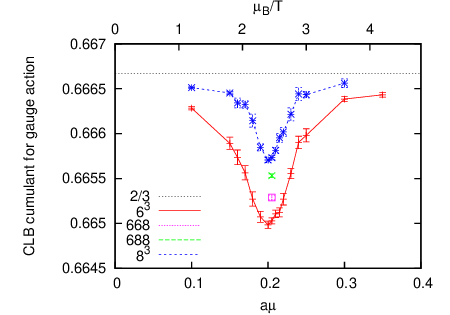}}
&
\scalebox{0.87}{\includegraphics{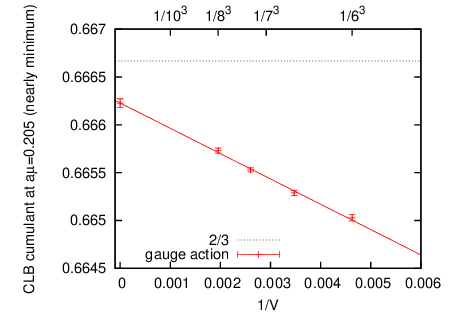}}
\end{tabular}
\end{center}
\caption{CLB cumulant of gauge action as function of $a\mu$ (Left) and
the volume scaling of minimum of the CLB cumulant (Right).
The curves in the right figure is a linear fit $\gamma+\delta/V$
where $\gamma$ and $\delta$ are fitting parameter.
}
\label{fig:gact_CLB}
\end{figure}

In Fig.\ref{fig:gact_CLB}, we show the Challa-Landau-Binder (CLB) cumulant \cite{Fukugita:1989yw,Challa:1986sk} of the gauge action $S_{\rm G}$ which is defined as
\be
U_G
=
1-\frac{\langle S_{\rm G}^4\rangle}{3\langle S_{\rm G}^2\rangle^2}.
\ee
When the minimum of the CLB cumulant in the thermodynamic limit is $2/3$, the transition is a crossover, while for other values it is of first order. 
The right panel in Fig.\ref{fig:gact_CLB} shows the volume dependence of the minimum.
The data is beautifully on the $1/V$ scaling line, and the thermodynamic limit value is different from $2/3$.
Therefore the CLB cumulant analysis indicates that the transition is consistent with being of first order. The kurtosis and the CLB cumulant show quite different behaviors.

\section{Distribution argument}

We now argue that the discrepancy in the scaling behavior observed above can be understood by a phenomenological distribution argument.
At a first-order transition point, the distribution of an observable $X$ can be approximately described by a double peak gaussian form given by 
\be
P(x)
\propto
e^{-\frac{(x-X_-)^2}{2c/V}}
+
e^{-\frac{(x-X_+)^2}{2c/V}}.
\ee
One can then calculate the susceptibility, the kurtosis and the CLB cumulant in terms of the parameter $c$, gap $\Delta=(X_+-X_-)/2$ and average $X=(X_++X_-)/2$, and find 
\bea
\chi_X
&=&
V
\langle
(x-\langle x\rangle)^2
\rangle
=
V\Delta^2+c,
\\
K_X
&=&
\frac{
\langle
(x-\langle x\rangle)^4
\rangle
}{
\langle
(x-\langle x\rangle)^2
\rangle^{2}
}
-3
=
\frac{
-2
}{
(1+\frac{c}{\Delta^2}\frac{1}{V})^2
}
=
-2
\left[
1-
\frac{2c}{\Delta^2}
\frac{1}{V}
+O(V^{-2})
\right],
\label{eqn:kurtosis}
\\
U_X
&=&
1-\frac{1}{3}
\frac{
\langle
x^4
\rangle
}{
\langle
x^2
\rangle^2
}
=
\frac{2}{3}
\frac{X^4+\Delta^4}{(X^2+\Delta^2)^2}
\frac{1}{\left(1+\frac{c}{X^2+\Delta^2}\frac{1}{V}\right)^2}
\nonumber\\
&=&
\frac{2}{3}
\frac{X^4+\Delta^4}{(X^2+\Delta^2)^2}
\left[
1-\frac{2c}{X^2+\Delta^2}\frac{1}{V}
+O(V^{-2})
\right],
\eea
where $\langle f(x)\rangle=k\int_{-\infty}^{\infty} dx f(x) P(x)$ with a normalization constant $k$.
From the above equations, we read that the peak of the susceptibility increases in proportional to the volume and the minimum of the kurtosis and the CLB cumulant have $1/V$ scaling as expected.
Note that such volume dependences originate from a non-zero gap $\Delta\neq0$ and $c\neq0$ which in turn  is nothing but the two-state signal of the first-order phase transition.
In the last section, we observed that the peak of the susceptibility shows a linear volume dependence.  Therefore we can say that $\Delta$ is non-zero and this is consistent with a first-order  phase transition.

Although the kurtosis and the CLB cumulant have a similar volume scaling behavior in the above equations, there is an important practical difference between them.
In the coefficient of $1/V$, there is $X$ in the latter but not in the former.
Depending on the size of $X$,  the parameter controlling the $1/V$ correction changes. 
For example, the CLB cumulant of the gauge action analyzed in the last section showed a nice scaling behavior.
This is because the average value is much larger than the gap $X\gg\Delta$ and the fluctuation is small $c\ll X^2$.
The histogram of the gauge action on the phase quenched configuration supports these relations.
On the other hand, the kurtosis showed a bad scaling behavior.
The main reason for this is as follows.
From the fitting the susceptibility we find  that $c\approx V\Delta^2$.  This suggests that the overlap of the two peaks is very large, which is roughly confirmed in the histogram on the phase quenched configurations.
In such a small volume, the $1/V$ expansion for kurtosis in eq.(\ref{eqn:kurtosis}) is not valid.  
Therefore the kurtosis does not have a good scaling behavior in contrast to the CLB cumulant.

\section{Summary}

We have observed that the susceptibility and the CLB cumulant have a nice scaling behavior, while the kurtosis does not.  We have argued that such a discrepancy is understood by the phenomenological double peak distribution analysis.  
This argument shows that even if the overlap of the two peaks are large (which often happens in a small volume setup),
the CLB cumulant and the susceptibility are still good indicators for the determination of the order of phase transition, especially for the CLB cumulant
under the condition that the average of the observable is larger than the gap of two peaks .
The presence of the average in the CLB cumulant makes $1/V$ scaling clearly visible.
Overall, we conclude that at this parameter set, the transition is consistent with being of first order. 

In order to check the above conclusion, we have performed further analyses.
For example, see Ref.~\cite{Jin} for an analysis of partition function zeros.  
Furthermore, a preliminary analysis for the 3-flavor theory is presented in Ref.~\cite{Nakamura}.


\vspace{3mm}
This work is supported in part by the Grants-in-Aid for
Scientific Research from the Ministry of Education, 
Culture, Sports, Science and Technology 
(Nos.
23105707, 
23740177, 
22244018, 
20105002). 
The numerical calculations were done on T2K-Tsukuba and HA-PACS cluster system
at University of Tsukuba.
We thank the Galileo Galilei Institute for Theoretical Physics for the hospitality and INFN for partial support offered to S.T. during the workshop ``New Frontiers in Lattice Gauge Theories", while this work was completed.

\end{document}